\nonstopmode

\documentclass[prb,showpacs,twocolumn]{revtex4}
\usepackage{graphicx}
\usepackage[cmex10]{amsmath}
\usepackage{amsthm, amssymb}
\usepackage{amsmath}
\usepackage{mdwmath}
\usepackage{mdwtab}

\def\gammat{\tilde\gamma}

\def\nt{\tilde n}

\def\jlt#1{{J. Lightwave Technol.} \textbf{#1}}

\def\opex{ Opt.\ Express }

\def\apl{ Appl.\ Phys.\ Lett.\ }

\def\josab{ J.\ Opt.\ Soc.\ Am.\ B }

\def\ol{ Opt.\ Lett.\ }

\begin{document}
\title{Saturable absorption in multi-core fiber couplers}
\author{Elham Nazemosadat and Arash Mafi}
\thanks{corresponding author.}
\affiliation{Department of Electrical Engineering and Computer Science, University of Wisconsin-Milwaukee, Milwaukee, WI 53211, USA}
%
\date{26 July 2013}

\begin{abstract}
The saturable absorption characteristics of two-, three-, and five-core 
one-dimensional fiber coupler arrays and the seven-core hexagonal 
fiber coupler array are investigated. It is shown that the 
performance of all these saturable absorbers are comparable and not 
much is gained, if anything, by going from a two-core nonlinear coupler 
geometry to a higher number of cores. This observation is supported by 
the similarity of the saturable absorption curves, as well as comparable pulse
characteristics obtained from the simulation of a generic mode-locked fiber laser cavity.
\end{abstract}

\pacs{42.81.Qb, 42.55.Wd, 42.65.Re, 42.65.Tg, 42.65.Jx, 42.65.Pc}
\maketitle

The operation of a mode-locked laser requires a form of intensity discrimination otherwise known as
saturable absorption~\cite{Garside}. Common saturable absorbers (SA) include semiconductor 
SA mirrors (SESAM)~\cite{Keller}, 
carbon nanotubes~\cite{Set}, graphene~\cite{Hasan}, Kerr lensing~\cite{Spence}, 
and nonlinear polarization rotation (NPR)~\cite{Noske}. Desired attributes of SAs include fast response time (relaxation time),
stability, long-term reliability, ease of use, appropriate saturation fluence, and low loss~\cite{Haiml}. 

For mode-locked fiber laser applications, fully integrable SAs are desired in order to take full
advantage of a robust alignment-free fiber cavity. Common fiber-integrable SA solutions have 
deficiencies that limit their usefulness for mode-locked fiber laser applications: SESAMs generally have 
picosecond response times, making them less desirable for ultrashort pulse generation, and their 
nonsaturable losses can result in excessive heating~\cite{Paschotta}. 
The long-term reliability of carbon nanotube- and graphene-based SAs when exposed to 
intense optical pulses is at best questionable~\cite{Song}, and both are limited in
modulation depths. SAs based on nonlinear polarization rotation are 
very sensitive to the slightest perturbation of the fiber cavity, making them unsuitable outside the laboratory. 

Another problem with the present SA technology concerns the rapid growth in the maximum achievable 
pulse energy and peak power in mode-locked fiber lasers 
over the past decade~\cite{Ortac,Lefrancois1,Lefrancois2}. SAs 
are becoming a limiting factor in scaling up the pulse energy and peak power to higher values. 
SESAMs, carbon nanutubes, and graphene will likely 
not be suitable for ultrahigh pulse energies due to heat generation and long-term reliability. 
The periodic SA curve in a nonlinear polarization rotation SA with respect to the pulse power 
and its sensitivity to environmental perturbations
makes this technique unsuitable for ultrahigh-peak-power mode-locked lasers. 

Nonlinear mode coupling in waveguide arrays (NMCWA), including multi-core fiber implementations, offers an attractive 
alternative that addresses most of the limitations faced by common SAs. 
A dual-core fiber laser geometry as an embedded SA was originally proposed by Winful et al.~\cite{Winful}. 
Since then, several theoretical studies have been conducted on nonlinear mode coupling 
in semiconductor coupled waveguide arrays~\cite{Eisenberg, Proctor, Hudson}, and 
multi-core optical fibers~\cite{Friberg,Betlej}. More recently, a seven-core tapered 
optical fiber was explored in the nonlinear regime and it was shown to have the potential 
to be used as an SA~\cite{Thomas}. 

A mode-locked fiber laser using an 
AlGaAs waveguide array SA was recently demonstrated experimentally~\cite{Chao}.
This demonstration clearly serves as a convincing experimental proof-of-concept for the
potential application of NMCWA as an SA. However, for practical device applications and improved efficiency,  
the insertion loss of the waveguide array needs to improve. 
An all-fiber SA in the spirit of Refs.~\cite{Winful,Friberg,Betlej} seems like
an ideal solution to the insertion loss problem, because it can be easily spliced to other components in the fiber cavity.
Although semiconductor arrays benefit from a larger nonlinear coefficient resulting in desirably short SAs,
fibers made from chalcogenide glasses with nonlinear coefficients of up to 1000 times larger than 
silica can provide similar performances~\cite{Lenz,Sugimoto}. In addition, the nearly instantaneous response of 
nonlinearities in optical fibers is ideal for ultrashort pulse generation~\cite{Sugimoto,AgrawalBook}.

\begin{figure}[htb]
\centering
\includegraphics[width=3.1in]{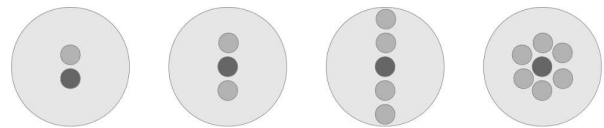}
\caption{The two-, three-, and five-core 1D arrays and the seven-core hexagonal array are shown in a 
fiber-optic geometry. All cores are assumed to be identical in a given geometry. The launch core
is identified with a darker shade.}
\label{fig:multiple-cores}
\end{figure}

In this paper, we study the performance of the multi-core SAs as a function of the number of the
waveguide cores. Many-core geometries have been the center of attention in NMCWA SAs; e.g.,
the NMCWA studied in Ref.~\cite{Proctor} has 41 coupled waveguides. 
We explore and compare the saturable absorption characteristics of two-, three-, and five-core 
one-dimensional (1D) arrays and the seven-core hexagonal array illustrated in Fig.~\ref{fig:multiple-cores}. 
The main question that is addressed in this paper is whether increasing the number of the 
cores improves the saturable absorption characteristics of these nonlinear multi-core arrays. 

The results indicate that the performance of all these SAs are comparable and not much is gained, if anything, by
going from a two-core nonlinear coupler geometry to a higher number of cores. 
This observation is supported by the similarity of the saturable absorption curves, as well as comparable
pulse characteristics obtained from the simulation of a generic mode-locked fiber laser cavity.
In the latter case, these SAs are placed in a fiber laser cavity, in which self-starting stable
mode-locked pulses are generated. According to the results, one can benefit from the simpler setup
of a two-core fiber with optimized parameters to get the desired output mode-locked pulses
instead of using the more complex multi-core fibers. 

In the following, for each multi-core fiber SA in Fig.~\ref{fig:multiple-cores}, the optical beam is launched in the {\em launch core} identified with 
a darker shade and is collected from the same core at the other end after propagating 
through the length of the SA. The dependence of the transmission through the multi-core fiber
on the optical power serves as the desired saturable absorption mechanism.  

\begin{figure}[htb]
\centering
\includegraphics[width=2.5in]{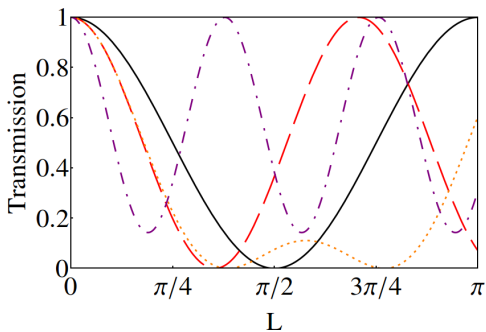}
\caption{The transmission through the two-core (solid black), 
three-core (dashed red), five-core (dotted orange) fiber 1D arrays and the seven-core 
hexagonal array (dash-dotted purple) is shown as a function of the normalized length 
in the linear regime.}
\label{fig:tau-z1}
\end{figure}

In the linear regime where optical power is low, neighboring waveguides exchange optical 
power periodically; the linear coupling is caused by the modal overlap of adjacent waveguides
and is most efficient when the adjacent modes have identical propagation constants~\cite{Saleh}. 
In Fig.~\ref{fig:tau-z1}, the transmission through the multi-core arrays of Fig.~\ref{fig:multiple-cores} 
is shown in the linear regime as a function of the normalized
length $L$ relative to ${\tilde c}^{-1}$; ${\tilde c}$ is the linear coupling coefficient between adjacent cores and 
is assumed to be the same among all neighboring cores in a given geometry.
The half-beat-length $L_c$ is defined as the distance over which the energy is maximally exchanged 
from the launch core to the neighboring cores and depends on the number of fiber cores and the geometry.

It can be seen that the half-beat-length of a three-core fiber is shorter than the two-core fiber,
because its launch core is coupled to ``more'' side cores with the same strength and can exchange
the optical power more efficiently. The linear dynamics in the case of five cores is a bit more 
complex due to the differing power exchange efficiencies of the side cores, but the first minimum 
happens for only a slightly longer SA device. In the case of the seven-core hexagonal array, the central core 
couples to six cores in the outer ring, resulting in a rapid exchange of optical power and shorter
half-beat-length; however, the power in the central core always remains larger than zero where the minimum
value of transmission is $1/7\approx 14.3\%$.  
In all these cases, the half-beat-length is inversely proportional to the coupling strength and is given by
$ {\tilde c}L_c=\xi_n (\pi/2)$, where $\xi_n$ is a numerical coefficient for the n-core scenarios studied here.
We have $\xi_2=1$, $\xi_3=1/\sqrt{2}\approx 0.71$, $\xi_5=4/\sqrt{27}\approx 0.77$, and $\xi_7\approx 0.378$.

In the nonlinear regime where optical power is high, self-phase and cross-phase modulation (SPM and XPM) effects 
alter the refractive index of each waveguide and consequently detune the effective propagation 
constants of the modes, reducing the power exchange efficiency between adjacent cores~\cite{Jensen}. In other words,
if high power is launched into a waveguide, the effective couplings to neighboring cores are reduced and 
the light remains mainly in the launch core. If the total length of the coupler in each case is chosen to be equal 
to the half-beat-length, for an optical pulse transmitting through the fiber array, its low intensity 
sides are efficiently channeled to the adjacent cores, while its high intensity center peak remains in the launch core, 
resulting in a power-dependent transmission and intensity discrimination~\cite{Winful,Proctor}. 
This saturable absorption feature can be used to produce the required pulse shaping for stable and robust 
mode-locked pulse trains.

The differential equation describing the propagation of the electric field in the $n$th core 
of a multi-core fiber nonlinear coupler, where only the interactions between the nearest 
neighbors are considered, is given by
\begin{align}
\label{eq:masterCoupling}
\dfrac{\partial E_n}{\partial \zeta}=i\sum_{\nt} E_{\nt}
+i\gammat_n|E_n|^2 E_n+i\sum_{\nt} {\mu_{n,\nt} |E_{\nt}|^2E_n}.
\end{align}
$E_n$ is the electric field envelope normalized to the pulse peak power $P_0$ at the entrance of the launch core.
The $\nt$ sums are over the cores adjacent to the $n$th core.
The propagation distance is rescaled to $\zeta={\tilde c}z$. 
$\gammat_n=\gamma^*_n P_0/{\tilde c}$ is a dimensionless parameter
characterizing the nonlinear strength, 
where $\gamma^*_n$ is the SPM coefficient of the $n$th core and $\mu_{n,\nt}$ 
is the cross-phase modulation coefficient between the $n$th core and the adjacent cores. 
The XPM terms are much smaller than the SPM terms and are 
thus neglected, i.e., $\mu_{n,\nt}\approx 0$. The dispersion and loss of the fiber coupler can be 
neglected if the SA length is sufficiently smaller than its dispersion length and effective length~\cite{AgrawalBook}.

\begin{figure}[htb]
\centering
\includegraphics[width=2.5in]{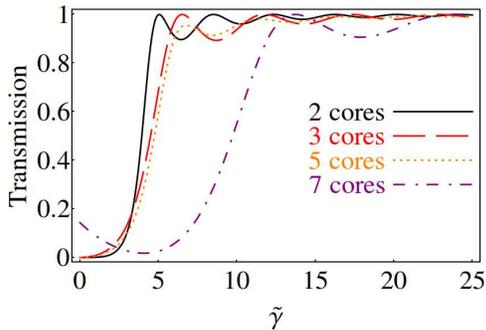}
\caption{Nonlinear transmission through a two-core (solid black), 
three-core (dashed red), five-core (dotted orange) fiber 1D arrays and the seven-core 
hexagonal array (dash-dotted purple) is shown as a function of the normalized nonlinear 
strength $\gammat$, which is assumed to be the same for all cores. The length of each 
SA array is chosen to be equal to the half-beat-length.}
\label{fig:Tau-vs-P0-case1}
\end{figure}

The nonlinear consequences of the coupled beam propagation in multi-core arrays have been
previously explored in detail; e.g., in Ref.~\cite{Christodoulides}.
Here, we borrow from these studies to obtain the saturable absorption curves of the multi-core
arrays of Fig.~\ref{fig:multiple-cores}. The length of each SA array is assumed to be equal to
the half-beat-length to ensure minimum transmission in the low-power linear regime.
The saturable absorption curves are shown in Fig.~\ref{fig:Tau-vs-P0-case1}
where the transmission through the launch core of the multi-core SA array is plotted  
as a function of the normalized nonlinear strength $\gammat$, which is assumed to be 
the same for all cores. The saturable absorption is dictated by a competition between 
the linear coupling term $i\sum_{\nt} E_{\nt}$ versus the nonlinear SPM term 
$i\gammat_n|E_n|^2 E_n$ in Eq.~\ref{eq:masterCoupling} and the transition from low 
transmission to high transmission occurs when these two terms are comparable. 

The three-core fiber SA has two adjacent cores coupled to its launch core
and consequently has a larger linear coupling term compared with the two-core scenario; 
therefore, more optical power (larger SPM term) is required to overcome the 
linear coupling term. That is why the three-core SA requires slightly higher power for nonlinear
saturation. 
The five-core and three-core scenarios are nearly identical, because even in the five-core
scenario, the launch core is only adjacent to two neighboring cores. The saturable absorption
curves for a higher number of cores in 1D linear arrays are practically identical to the five-core SA. 

\begin{figure}[htb]
\centering
\includegraphics[width=2.5in]{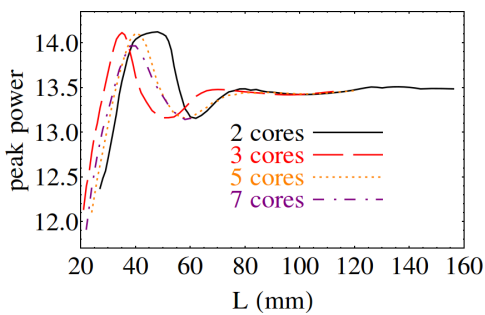}
\caption{The normalized peak power of the mode-locked pulses are shown as a function of the 
length the SA, where the core-to-core coupling coefficients $c$ are adjusted so that the length of the SA is
equal to the half-beat-length.}
\label{fig:Pout-vs-Length}
\end{figure}

For the seven-core hexagonal array SA, the nonlinear SPM term must overcome the linear coupling to 
six adjacent cores and requires higher normalized nonlinearity to saturate the SA. However, the 
SA curve closely resembles those of 1D arrays and can be mapped to the 1D saturable absorption curves
by adjusting the core-to-core ${\tilde c}$ coefficient. In fact, if SA devices are chosen at the same 
length, where ${\tilde c}$ must inevitably be smaller for the seven-core array, the saturable 
absorption curves for all the geometries in Fig.~\ref{fig:multiple-cores} will be nearly identical
as a function of the input power. This can easily be seen if the ``horizontal scale'' for the seven-core
SA curve in Fig.~\ref{fig:multiple-cores} is rescaled by a factor of approximately $2/6$ to compensate
for the six adjacent cores versus the two adjacent cores in the three-core fiber. The SA curves nearly
overlap after such a rescaling. This fact will result in nearly identical pulse characteristics
in mode-locked cavity simulations of Fig.~\ref{fig:Pout-vs-Length}, as will be discussed shortly. 

There is another notable difference in the seven-core saturable absorber: the transmission initially
drops to zero at finite $\gammat$ before saturating at higher power. This effect is due to the 
complex coupling pattern in the seven-core geometry. A nearly monotonic increase can be artificially
created by shortening the fiber below the half-beat-length at the expense of a smaller modulation depth. 
  
To study the mode-locking behavior of the multi-core array, it is placed within the mode-locking 
optical cavity. A typical passive mode-locked fiber laser consists of an active single mode fiber with a
bandwidth-limited gain, which compensates for the energy loss in the cavity, and an
intensity discrimination device as an SA. When the interactions of the attenuation, chromatic dispersion, nonlinearity,
and bandwidth-limited gain in the optical fiber cavity are combined with the intensity discrimination, a train of
mode-locked pulses is formed from white noise after several numbers of round trips in the cavity. These interactions
are captured by two sets of equations: one is the nonlinear Schr\"{o}dinger equation governing the pulse propagation
in the cavity, while the other is Eq.~\ref{eq:masterCoupling} describing the saturable absorption.

The pulse propagation in the cavity is given by~\cite{Haus}
\begin{align}
\label{eq:masterNL}
\dfrac{\partial A}{\partial z}=-\alpha A+
\dfrac{i}{2} \dfrac{\partial^2 A}{\partial T^2}+i\eta|A|^2 A+g(z)(1+\tau \dfrac{\partial^2}{\partial T^2})A,
\end{align}
where $A(z,T)$ is the normalized electric field envelope, $z$ is the propagation distance along the fiber, and 
$T$ represents the time measured in the retarded frame of the pulse normalized by a constant $T_0$, which is the expected
pulse duration. 
$A(z,T)$ is normalized by the peak power of the fundamental 
soliton $P_{fs}=|\beta_2|/(\gamma T_0^2)$, where $\beta_2$ is the group velocity dispersion of the fiber, and
$\gamma$ is the fiber nonlinear coefficient~\cite{AgrawalBook}. 

It is also implicitly assumed 
that the wavelength of the pulse $\lambda_0=1.55~\mu m$ 
falls within the anomalous-dispersion regime of the fiber. The propagation distance $z$ is scaled to the dispersion length $L_D=T_0^2/|\beta_2|$, 
and ${\eta}$ is the dimensionless cavity nonlinear coefficient given by ${\eta}=\gamma P_{fs} L_D$.
The normalized attenuation is considered through $\alpha$, while the bandwidth-limited gain of the cavity is described 
by the parameters $g(z)$ and $\tau$. The dimensionless parameter $\tau$ controls the pulse width in the mode-locking 
process and is given by $\tau=1/(\Delta\omega T_0)^2$ where $\Delta\omega=(2\pi c/\lambda_0^2)\Delta\lambda$ is the 
spectral gain bandwidth and $\Delta\lambda$ is the gain bandwidth~\cite{Proctor2}. The gain saturation $g(z)$ is given by
\begin{align}
\label{eq:gz}
g(z)=\dfrac{2g_0}{1+\|A\|^2/ e_0},
\end{align}
where $g_0$ and $e_0$ are the pumping strength and the cavity saturation energy, respectively, 
and $\|A\|^2=\int_{-\infty}^\infty \mathrm{|A|}^{2}\,\mathrm{d}T$ is the pulse energy.

The simulations are based on Eq.~\ref{eq:masterNL}, while Eq.~\ref{eq:masterCoupling} 
is applied in every round trip of the laser cavity. The coupling efficiency of the cavity to the
launch core of the multi-core SA is assumed to be 100\% on both the entrance and the exit ports; therefore, 
$A$ in Eq.~\ref{eq:masterNL} becomes $E$ at the entrance in the launch core of the multi-core geometry in 
Eq.~\ref{eq:masterCoupling} and vice versa at the exit.

For the mode-locking cavity simulations, the initial pulse conditions are assumed to be 
low amplitude noise fluctuations. The cavity length is 5m, while the rest of the parameters 
used are $T_0=114$~fs, $\alpha=0.05$, $\Delta\lambda=35~nm$, $g_0=0.39$ and $e_0=1$~\cite{Proctor2}.
We also have $\beta_2=-0.0153~ps^2/m$ resulting in $L_D=0.84$~m. The nonlinear coefficient of the cavity is chosen 
as $\gamma=1.7~{\rm km^{-1}W^{-1}}$. The nonlinear coefficient of the SA is assumed to be
$\gamma^*=0.34~{\rm m^{-1}W^{-1}}$, which can be obtained by using a chalcogenide glass fiber~\cite{Lenz}.

Fig.~\ref{fig:Pout-vs-Length} shows the normalized power of the mode-locked pulses after 100 
round trips in the laser cavity, resulting in full convergence, as a function of the length the SA. 
For each length (and for each SA), the core-to-core coupling coefficients ${\tilde c}$ are adjusted so that the length of the SA is
equal to the half-beat-length for optimum performance. This choice of the coupling coefficients 
ensures that the power is maximally transferred from the launch core to the other cores in the 
linear regime. In practice, the core-to-core coupling coefficients ${\tilde c}$ can be designed and controlled
via core-to-core separations. 
As we discussed above, the saturable absorption curves for all multi-core arrays studied here 
are nearly identical when the SAs are of the same length. Therefore, it should come at no surprise that
the normalized pulse peak power is nearly the same for all these SAs as shown in Fig.~\ref{fig:Pout-vs-Length}.   

In summary, the results in Figs.~\ref{fig:Tau-vs-P0-case1} and~\ref{fig:Pout-vs-Length} show that nearly identical
performance is expected from the geometries that we explored. While that results are seemingly generic 
for popular SA geometries, we emphasize that this study is not exhaustive.  
An infinite number of multi-core geometries can be constructed even with non-uniform core-to-core couplings
and non-identical core sizes applied to many different choices of laser parameters. 
Therefore, the main conclusion one can draw from this study is that when an NMCWA is to be used as an SA, 
it is important to verify whether multiple cores yield any tangible performance benefits besides making 
the device unnecessarily more complex. Future efforts will focus on comparisons between multi-mode and multi-core
all-fiber SA devices~\cite{Elham}.
\section*{Acknowledgments}
The authors acknowledge support from the Air Force Office of Scientific Research under Grant FA9550-12-1-0329.

\end{document}